\documentclass[conference]{IEEEtran}
\usepackage{cite}
\usepackage{tikz}
\newcommand\copyrighttext{%
  \footnotesize \textcopyright 2017 IEEE. Personal use of this material is permitted. Permission from IEEE must be obtained for all other uses, in any current or future media, including reprinting/republishing this material for advertising or promotional purposes, creating new collective works, for resale or redistribution to servers or lists, or reuse of any copyrighted component of this work in other works.}
\newcommand\copyrightnotice{%
\begin{tikzpicture}[remember picture,overlay]
\node[anchor=south,yshift=12pt] at (current page.south) {\fbox{\parbox{\dimexpr\textwidth-\fboxsep-\fboxrule\relax}{\copyrighttext}}};
\end{tikzpicture}%
}

\usepackage[outdir=./]{epstopdf}
\usepackage{graphicx}
\graphicspath{{../}}
\DeclareGraphicsExtensions{.eps}

\usepackage{amsmath}
\usepackage{algorithmic}
\usepackage{caption}
\ifCLASSOPTIONcompsoc
 \usepackage[caption=false,font=normalsize,labelfont=sf,textfont=sf]{subfig}
\else
 \usepackage[caption=false,font=footnotesize]{subfig}
\fi

\usepackage{color}
\usepackage{booktabs}
\usepackage{tabularx}
\usepackage{multirow}
\usepackage{arydshln}
\newcolumntype{P}[1]{>{\centering\arraybackslash}p{#1}}
\newcolumntype{M}[1]{>{\centering\arraybackslash}m{#1}}

\begin{document}
\addtolength{\parskip}{-0.5mm}

\title{Measuring the Eccentricity of Items}

\author{\IEEEauthorblockN{Chanyoung Park}
\IEEEauthorblockA{School of Integrated Technology\\
Yonsei University\\
Incheon, Korea\\
Email: chanpark@yonsei.ac.kr}
\and
\IEEEauthorblockN{Songkuk Kim}
\IEEEauthorblockA{School of Integrated Technology\\
Yonsei University\\
Incheon, Korea\\
Email: songkuk@yonsei.ac.kr}
}

\maketitle
\copyrightnotice

\begin{abstract}
The long-tail phenomenon tells us that there are many items in the tail. However, not all tail items are the same. Each item acquires different kinds of users. Some items are loved by the general public, while some items are consumed by eccentric fans. In this paper, we propose a novel metric, \textit{item eccentricity}, to incorporate this difference between consumers of the items. Eccentric items are defined as items that are consumed by eccentric users.
We used this metric to analyze two real-world datasets of music and movies and observed the characteristics of items in terms of eccentricity.
The results showed that our defined eccentricity of an item does not change much over time, and classified eccentric and noneccentric items present significantly distinct characteristics.
The proposed metric effectively separates the eccentric and noneccentric items mixed in the tail, which could not be done with the previous measures, which only consider the popularity of items.
\end{abstract}
\IEEEpeerreviewmaketitle

\section{Introduction}
It is widely acknowledged that items in various markets follow a long-tailed distribution.
A long-tailed distribution suggests the importance of tail items, which in aggregate make up a significant portion of the total market.
Such less-popular items are important not only because of their market size but also because of the information they give about the users. Every person consumes their own special items, and these eccentric ones may imply more about the users' preference than other popular items.

However, not every less-popular item is special.
The tail of the market has many items, and they are not all the same.
Each item acquires different kinds of users.
Even with the same number of users, items may have different types of users.
Some items are loved by the general public, while some are mostly consumed by eccentric fans.
This difference in consuming users is more prominent in less-popular items because they do not have as many users as popular items do, and are more likely to appeal to a certain group of users.
Despite this intuitive difference in consumers between less-popular items, there have not been many studies that explored and captured this characteristic.

The importance of understanding unpopular items has been widely recognized by the field of recommender systems (RS).
Novelty has been regarded as an important aspect of RS because people seek not only relevant but also unexpected recommendations.
Several user studies have shown that a small increase in accuracy does not necessarily lead to a higher user satisfaction \cite{Konstan, McNee}, while serendipity notably influences user contentment.
Therefore, many approaches have been proposed to make novel recommendations \cite{Onuma, Kapoor}.

In the previous studies on the serendipity of RS, a trade-off between accuracy and diversity has been reported \cite{Park, Ge, Adomavicius, Ziegler}.
To mitigate the accuracy loss of incorporating serendipity, various methods to manage tail items have been proposed.
As one of the successful approaches, Park et al. \cite{Park} have clustered tail items into several groups and performed a regression within the group.
The improvement made by clustering tail items has implied that there are different kinds of items mixed in the long tail.
This supports our earlier explanation on the less-popular items.

In this paper, we therefore propose a new metric of item characteristics called \textit{item eccentricity} rather than a simple popularity represented by the number of feedbacks an item has.
We assumed that the core difference between the tail items lies in who are the consuming users.
An item is called rare when a small number of people have consumed the item. Eccentric items have something more. Their users must be eccentric too.
The coconsumer's usual behavior, like whether they are following mainstream or being an enthusiast of a niche genre, implies the degree of an item's eccentricity.
The \textit{item eccentricity} we propose considers the characteristics of users, and captures the significant difference between two similar terms, \textit{rarity} and \textit{eccentricity}.

Accordingly, we first start by defining the users' personal long-term degree of preference on rare items and refer to it as \textit{user eccentricity}.
We then define the \textit{item eccentricity} as the weighted average of user eccentricity (Section \ref{sec:def}).
Using the proposed metrics, we analyze the two real-world datasets, \textit{Last.fm} and \textit{MovieLens}, to demonstrate that our metric effectively separates the different tail items in terms of eccentricity (Section \ref{sec:analysis}).
We investigate whether the eccentricity of items is inherent and therefore maintained regardless of time, and see what is different between the eccentric and noneccentric items.
Finally, we conclude in Section \ref{sec:con}.

\section{Related Work}
In this section, we briefly review related works. We first explain how other studies have worked to incorporate serendipity in recommendations and why a metric to measure such a characteristic is needed.
Afterward, we investigate the previous approaches of defining and measuring the item eccentricity.

\textbf{Popularity Bias and Serendipitous Recommendation}
Many studies have shown that the item popularity in various markets follows a long-tailed distribution \cite{Celma, Goel, Sastry}.
Items in the head dominate the interaction between users and items, while a large number of items are located in the tail.
People are more likely to consume and produce feedback on the head items than on the tail items.
We call this bias to popular items \textit{popularity bias} \cite{Steck}.
Many RSs also tend to recommend popular items over unpopular items because they have enough ratings to be used by recommender algorithms, thus they can make a more accurate recommendation.
Several researchers have demonstrated the trade-off between accuracy and serendipity in the recommendation \cite{Ge, Adomavicius, Steck}.

Despite the relatively low performance on tail items, the importance of serendipity in RS has been shown by several user studies \cite{Tacchini, Zhang, McNee, Lee}.
In the studies, users are reported to seek both surprising and interesting recommendations, which we call serendipity \cite{Herlocker}.
The less-popular items may take up a small part of the overall interaction between users and items. However, each person still consumes their own unpopular but unique items for a significant proportion.
To reduce the decrease in accuracy of including serendipity in RS, various methods have been proposed.
Steck \cite{Steck} used a popularity-stratified training method to correct for popularity bias in the dataset.
Lu \cite{Lu} proposed a serendipitous personalized ranking model, which optimizes both accuracy and serendipity in a single learning procedure.
Park and Tuzhilin \cite{Park} grouped tail items using clustering methods and improved the recommendation accuracy.
The above works imply that there are groups of different items that are mixed in the tail part, and we need different ranking systems and methods to incorporate the differences between items.
Besides the simple popularity, the need for defining the new metric that shows a different perspective of items has been recognized.

\textbf{Item Eccentricity} There have been many approaches to measuring the extent to which the recommendations are serendipitous \cite{Hurley, Vargas}.
However, the metric on the item novelty has a relatively small number of studies, although it is important in understanding tail items.
Nakatsuji et al. \cite{Nakatsuji} proposed the Random Walk and Restart method over the user graph, and suggested how to find the novel items from relevant users, yet did not quantitatively measure the \textit{novelty}.
Marques et al. \cite{Marques} saw the novelty of items as having two other aspects besides being unexplored by a user: familiarity and mainstreamness. They defined the mainstreamness of music items as the log of the overall popularity measured by total play count.
Similarly, Vargas \cite{Vargas} defined item novelty as the log of the inverse popularity.
These metrics regarded every item with few users as novel and a deviation from the mainstream.
However, not every unpopular item is an eccentric item. Even with the same number of users, items may have different types of users.
The two terms, rarity and eccentricity, are often used interchangeably, yet they are different because the core concept of \textit{niche} or \textit{eccentric} depends on who are the coconsumers.
Therefore, to measure the eccentricity of items, a metric that takes characteristics of interacted users into account is required.

\begin{figure}[!t]
\centering
\includegraphics[width=0.98\columnwidth]{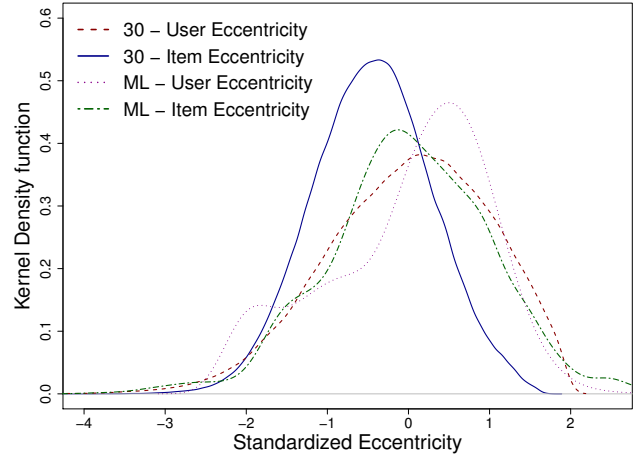}%
\label{fig:30_density}
\caption{Kernel density function of eccentricity metrics in the two datasets. 30 and ML in the legend denotes the \textit{30Music} and MovieLens datasets, respectively.}
\label{fig:density}
\end{figure}

\section{Definition}
\label{sec:def}

In this section, we present the definition of \textit{item eccentricity}. We first introduce the \textit{item rarity}, which represents the popularity of items as the well-known IDF term. This is similar to the previous works \cite{Vargas, Marques}, but we use this term to indicate the rarity, not the novelty of items.
We then define the \textit{user eccentricity} as a user's long-term tendency toward tail items.
We define it by how much a user consumes unpopular items and how much they like the unpopular ones.
This assumes that users who consistently consume and give positive ratings on unpopular items are eccentric.
Finally, we propose the \textit{item eccentricity} as the average of \textit{user eccentricity} of consumers weighted by their ratings.
We elaborate the definition of the three metrics in the following subsections.

\subsection{Item Rarity}
We assume that there is a set of users $U = {u_1, ..., u_N}$, a set of items $I = {i_1, ..., i_M}$, a set of time windows $T = {t_1, ..., t_L}$, and the set of known feedbacks $F = {f_{1,1,1}, ..., f_{u,i,t}, ..., f_{N,M,L}}$ produced by the customers in $U$ for the items in $I$ in the time windows in $T$.
Time windows are set to reflect the rise and fall nature of item popularity. Many items become less popular after the release and consuming the items in a different period should be treated differently in later processes.
The size of the time window can be determined differently depending on which dataset is being used.
We conducted preliminary research to find out which size provides the highest stability of popularity within the time window and moderate variability between the time windows.
Our results demonstrated that four weeks was the most suitable for the window size among the choices 1,2,3,4,6,12 weeks.
Therefore, in the rest of our analysis, we set the time windows to four weeks, where we approximately refer to it as one month.
The feedbacks can be either explicit, like five-scale ratings, or implicit, such as the number of plays on a track.

With explained notations, the rarity of an item $i$ in time window $t$ can be defined as the negative log of the number of feedbacks it has:
\begin{equation} \label{eq:ir}
Item\,\,Rarity_{i,t}\,(IR_{i,t}) = -log\,|F_{i,t}|,
\end{equation}
where $|F_{i,t}|$ indicates the number of feedbacks on item $i$ in time window $t$.
In the definition, we used the number of feedbacks on each item rather than a sum or average of feedbacks because the number of users is more directly related to the popularity.
After applying equation \ref{eq:ir}, we standardized the resulting values to a $z$-score. Unless otherwise noted, we use this standardized value when we speak of the \textit{item rarity} for the remainder of the paper.
When we refer to one representative rarity value of an item, $IR_i$, it is the average value of the rarity over all time windows of the item.

\subsection{User Eccentricity}
Eccentric users are characterized by their long-term tendency toward rare items.
It can be measured by two factors: degree of consuming unpopular items over popular items and preference on such items presented by user's feedback.
On each user's item list, the degree of consuming unpopular items can be measured as an average rarity of the items and use the feedback of the user to give the weight on each item's rarity:
% It can be measured by two factors: number of consumed unpopular items and preference on such items presented by user's feedback.
% On each user's item list, the number of used rare items can be measured as an average rarity of the items and use the feedback of the user to give the weight on each item's rarity:
\begin{equation} \label{eq:ue}
User\,\,Eccentricity_u\,(UE_u) = \frac{\sum_{i\in I_u,t\in T}f_{u,i,t}\cdot IR_{i,t}}{\sum_{i\in I_u,t\in T} f_{u,i,t}}
\end{equation},
where $I_u$ indicates the set of items that user $u$ has consumed and the $IR_{i,t}$ is the item rarity of item $i$ in time window $t$.
The value of user eccentricity is also standardized to a $z$-score as the item rarity was.
Users with higher eccentricity consume a greater number of rare items and give more positive feedback on such items.

\subsection{Item Eccentricity}
Based on the user eccentricity measured, we can finally define the item eccentricity.
We can express the item eccentricity as how much an item is consumed and liked by eccentric users, which is an average user eccentricity weighted by the feedback on the items:
\begin{equation} \label{eq:ie}
Item\,\,Eccentricity_i\,(IE_i) = \frac{\sum_{u\in U_i}f_{u,i}\cdot UE_{u}}{\sum_{u\in U_i} f_{u,i}}
\end{equation}.
$U_i$ is the set of users who have consumed the item $i$ and the $UE_{u}$ is the measured eccentricity of user $u$.
Item eccentricity is also standardized as other previous metrics were.

\section{Analysis on Eccentric Items}
\label{sec:analysis}
In this section, we present the analysis results on two real-world datasets using the proposed metrics.
Through the analysis, we show the stability of item eccentricity and the difference between eccentric and noneccentric items to demonstrate our metrics' validity.

\subsection{Datasets}
In our analysis, we use two datasets of different domains: the \textit{30Music} dataset \cite{Turrin} and the MovieLens 20M dataset. Table \ref{tab:data} describes the detailed statistics of the datasets.

\textbf{\textit{30Music} Dataset}
A publicly available dataset of user behavior logs collected from \textit{Last.fm}, an online music streaming service. It provides data about users, tracks, and the occurred music listening events as tuples of the form \texttt{<user, track, timestamp>}. It also provides a small amount of \textit{loved track} information collected from users.

\textbf{MovieLens 20M}
A user-movie ratings dataset collected from MovieLens.
It represents each rating tuple of the form \texttt{<user, movie, rating, timestamp>}.

The \textit{30Music} dataset exhibits a greater number of items and a more extreme long tail.
Another difference between the two datasets is the type of feedback given by users.
MovieLens provides explicit feedbacks as 0.5--5 rating while \textit{Last.fm} only provides implicit feedbacks like listening events.
As we need a set of numeric feedbacks $F$ in our metric definitions, for the \textit{30Music}, we define the $f_{u,i,t}$ as the number of times the user $u$ listened to the track $i$ in the time window $t$.

\begin{figure*}[!t]
\centering
\subfloat[30Music Dataset]{\includegraphics[width=0.98\columnwidth]{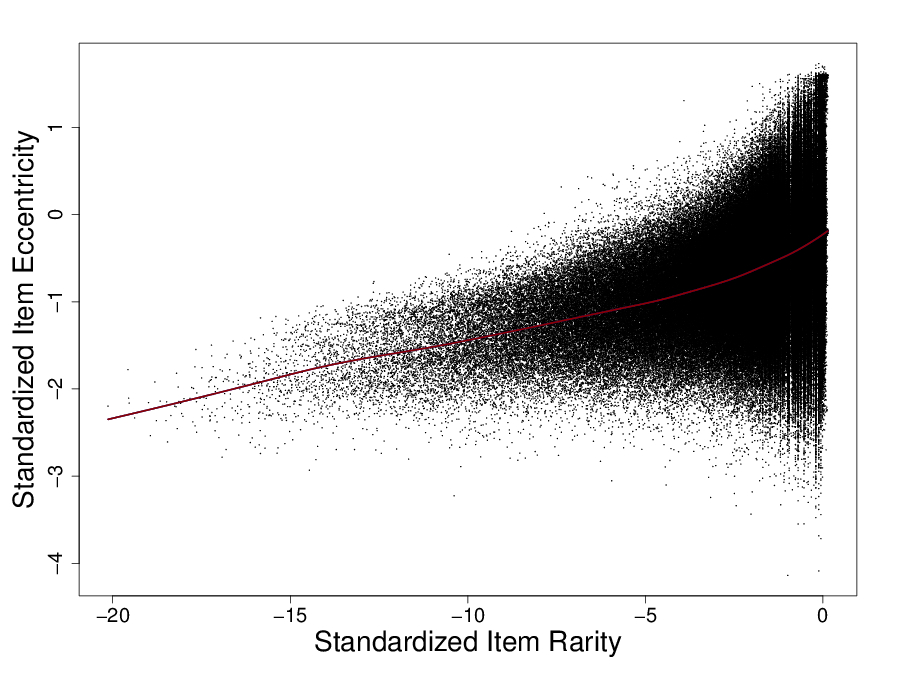}%
\label{fig:30_ire}}
\hfil
\subfloat[MovieLens Dataset]{\includegraphics[width=0.98\columnwidth]{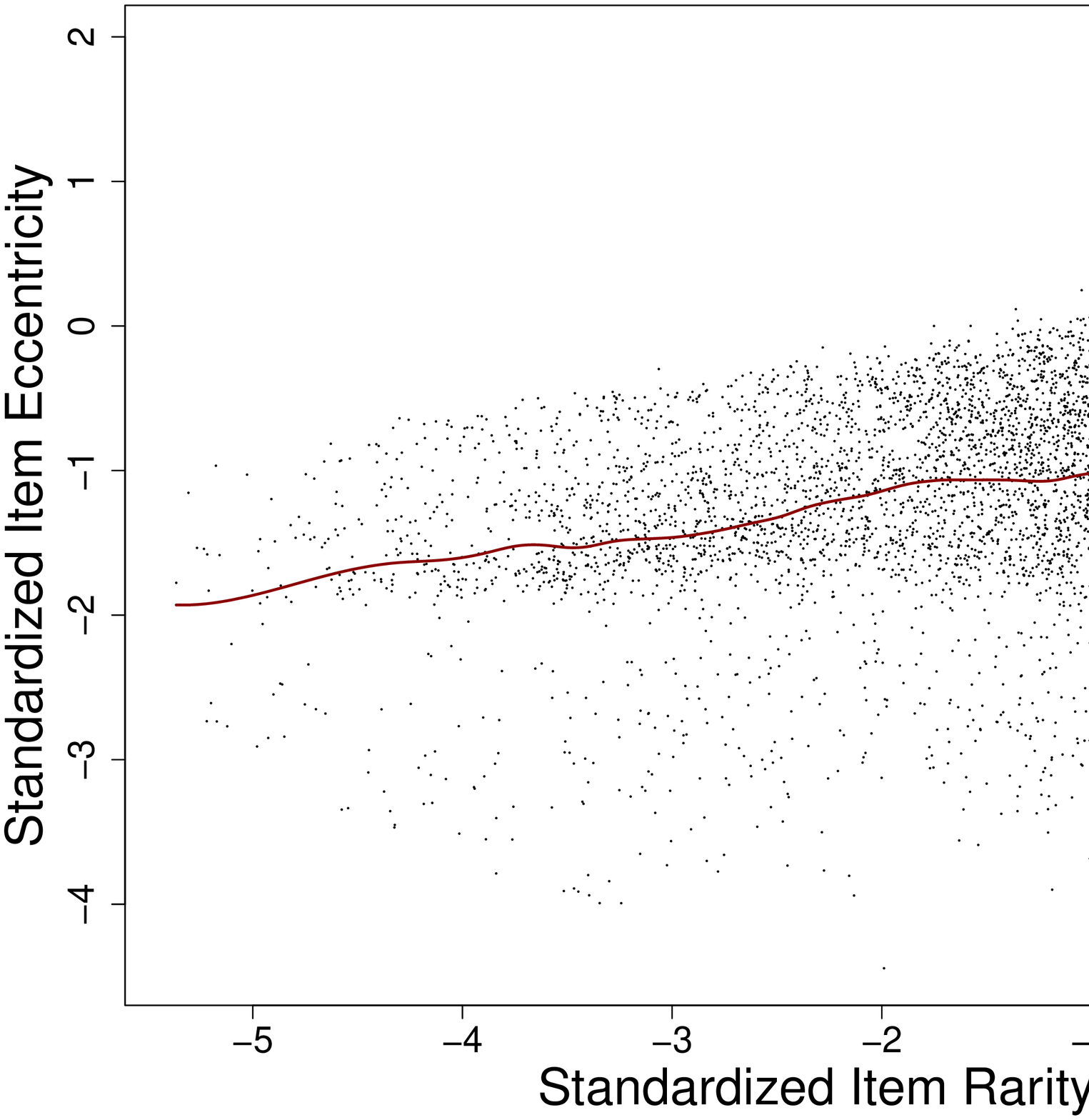}%
\label{fig:ml_ire}}
\caption{Graph of item eccentricity against item rarity. The red line indicates the regression line of the points.}
\label{fig:ire}
\end{figure*}

\begin{table}[t]
\centering
\caption{Statistical information of the two datasets. Row corresponding to feedback density shows the data density of the user-item matrix.}
\label{tab:data}
\resizebox{0.75\columnwidth }{0.18\columnwidth}{
\begin{tabular}{ccc}
\toprule
&\textit{30Music}& MovieLens \\
\midrule
the number of logs & 31.4M & 20M \\
$|U|$ & 45K & 138K \\
$|I|$ & 5.7M & 27K \\
value range of $f$ & 1-1504 & 0.5-5 \\
average value of $f$ & 1.45& 3,53\\
feedback density(\%) & 0.009& 0.540\\
\bottomrule
\end{tabular}}
\end{table}

\subsection{Distribution of Eccentricity}
In our analysis, we excluded items whose rarity value is zero for every time window.
These items have only one user for each time window and are not suitable to be analyzed for eccentricity because such items do not have coconsumers and solely depend on one user's eccentricity.
Figure \ref{fig:density} shows the kernel density of standardized user eccentricity and item eccentricity in the two datasets.
It presents the number of items and users on each eccentricity value.
All four eccentricity metrics from the two datasets have the same mean of zero, and the same standard deviation of one because all of them are standardized $z$-scores.

Because the user eccentricity is an average of item rarity weighted by the feedback, the user eccentricity does not follow the long-tail distribution as the item rarity does. It is relatively equally distributed over a wide range, which shows that users have different degrees of preferring the tail items.
The item eccentricity is the average user eccentricity weighted by users' feedback.
A shift of mode to lower value in item eccentricity illustrates the popularity bias explained earlier.
People tend to give better feedback on noneccentric items, and the most general eccentricity of the items becomes less eccentric compared with user eccentricity.
Therefore, the item eccentricity becomes more concentrated on the less eccentric region.
However, the degree of concentration on mode differed by the dataset.
Items in the \textit{30Music} dataset were more concentrated on the mode compared with the items in the MovieLens.

\subsection{Rarity and Eccentricity}
In this subsection, we compare the rarity and eccentricity of items and determine whether the eccentric and noneccentric items are genuinely different.

People do not make sharp distinctions between rarity and eccentricity in the real world. While they do seem to be similar to each other, our research suggests that the two metrics are not the same.
The most apparent distinction between rarity and eccentricity would be the distribution. The item rarity follows the long-tailed distribution, and the eccentricity follows the bell-curved distribution.
From the distribution, we can see that most items are rare, but only some items are eccentric.

\begin{figure*}[!t]
\centering
\subfloat[\textit{30Music} Dataset]{\includegraphics[width=0.85\columnwidth]{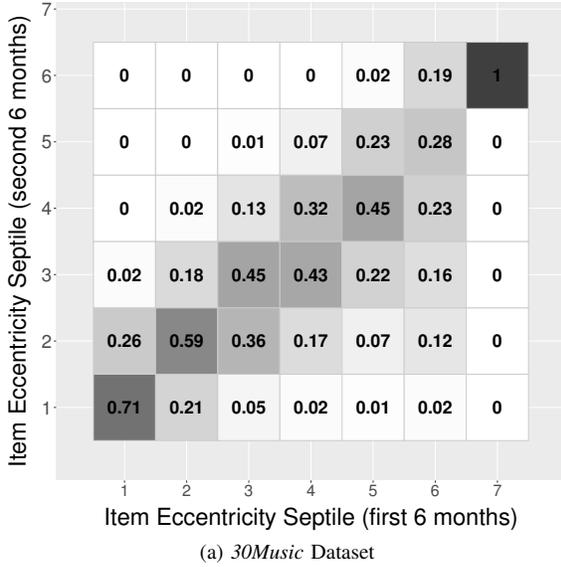}%
\label{fig:conf}}
\hfil
\subfloat[MovieLens Dataset]{\includegraphics[width=0.85\columnwidth]{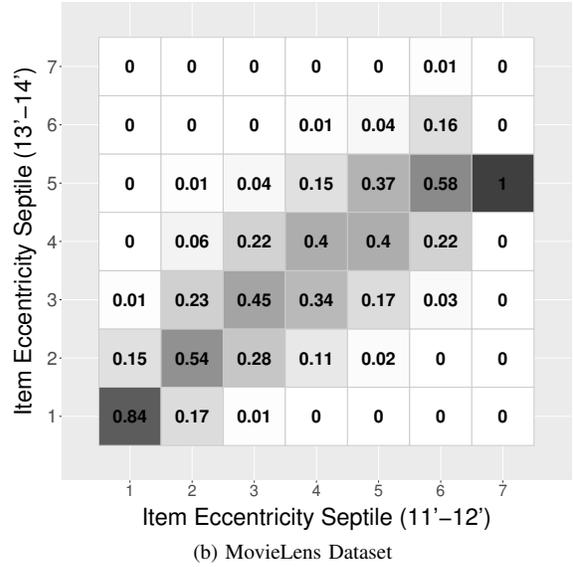}%
\label{fig:ml_conf}}
\caption{Probability table of item eccentricity measured in the first period changing to the one measured in the second period. Higher group number indicates higher eccentricity.}
\label{fig:conf}
\end{figure*}

Figure \ref{fig:ire} shows the item eccentricity versus rarity. Each point in the figure represents an item in the dataset.
We only plotted the items that have feedbacks for more than four time windows for stability. Therefore, the extremely unpopular items are excluded in the graph.
In the figure, we can observe that as items become unpopular, their eccentricity value diverges. It shows that there are items that have the same rarities but different eccentricities.
Considering that the item eccentricity is the weighted average of user eccentricity, this is a natural result of the central limit theorem.

Because unpopular items have a small number of users and their average eccentricity becomes the items' eccentricity, it is possible that high eccentricity is due to randomness, not due to the items' inherent characteristics.
If some unpopular items are achieving high or low eccentricity just by chance, they would not necessarily appear to be eccentric in every period. On the other hand, if one item is consistently consumed by eccentric users and thus really is eccentric, it should exhibit a high eccentricity value regardless of the time being measured.
Therefore, to demonstrate that eccentric items identified by our metric are genuinely eccentric, we examine item eccentricities of two separated periods and compare the two results.

First, we divided each dataset in two according to its timestamp.
For the MovieLens dataset, we got two datasets from 2011--2012 and 2013--2014. For the \textit{30Music} dataset, we got two datasets from February--July and August--January.
Within the separated datasets, we calculated the standardized item eccentricities.
We then divided items into seven septile groups by their eccentricity.
We then kept the items that have been consumed by more than 10 users in both periods, and that have not been released during the test period to avoid including items that have a too different condition in the two periods.

Figure \ref{fig:conf} shows the relationship between the two eccentricities at different times on the same item.
Each number in the box shows the probability of the septile group pair.
If the item eccentricity is obtained randomly, the probability of moving from one septile to other septiles in a different period should be the same across the groups.
However, in Figure \ref{fig:conf}, items that had high eccentricity in the first period had a higher possibility of having high eccentricity in the second period.
This shows the consistency of the item eccentricity, and proves that there are items that are naturally favored by eccentric users and items that are not.

\subsection{Comparison Between Tail-Item Groups}

\begin{table}[!t]
\renewcommand{\arraystretch}{1.1}
\caption{Average feature value of groups separated by rarity and eccentricity in the \textit{30Music} dataset. All pairs showed a significant difference (p$<$0.01).}
\label{tab:30group}
\centering
\resizebox{\columnwidth}{0.35\columnwidth}{
\begin{tabular}{m{4mm}M{11mm}M{12mm}M{8mm}M{10mm}M{10mm}}
\toprule
IR& IE Group & listened to\newline once users & \# of plays & \# of \newline loved users& \# of user \newline of artists \\
\midrule
\multirow{2}{*}{0.6} & Ecc. & 69\% & 2.11  & 1.03& 241\\
\cdashline{2-6}[0.5pt/1pt]
& Non-ecc.& 73\% & 1.83 & 0.66& 957 \\
\cmidrule(lr){2-6}
\multirow{2}{*}{0.7} & Ecc. & 68\% & 1.83 & 0.42& 145 \\
\cdashline{2-6}[0.5pt/1pt]
& Non-ecc.& 74\% & 1.64& 0.29 & 897 \\
\cmidrule(lr){2-6}
\multirow{2}{*}{0.8} & Ecc. & 66\% & 2.21 & 0.60& 188 \\
\cdashline{2-6}[0.5pt/1pt]
& Non-ecc.& 71\% & 1.83  & 0.39 & 844\\
\cmidrule(lr){2-6}
\multirow{2}{*}{0.9} & Ecc. & 66\% & 2.11  & 0.58 & 169\\
\cdashline{2-6}[0.5pt/1pt]
& Non-ecc.& 72\% & 1.83 & 0.38 & 828 \\
\cmidrule(lr){2-6}
\multirow{2}{*}{0.99} & Ecc. & 62\% & 2.11  & 0.73& 177\\
\cdashline{2-6}[0.5pt/1pt]
& Non-ecc.& 70\% & 1.83  & 0.52& 839\\
\bottomrule
\end{tabular}}
\end{table}

\begin{table}[!t]
\renewcommand{\arraystretch}{1.1}
\caption{Statistics of movies separated by rarity and eccentricity in the MovieLens dataset. The asterisk (*) indicates a significant difference (p$<$0.01).}
\label{tab:mlgroup}
\centering
\resizebox{\columnwidth }{0.35\columnwidth}{
\begin{tabular}{m{4mm}M{11mm}M{8mm}M{12mm}M{14mm}}
\toprule
IR& IE Group & rating & multimodality \newline of ratings & first two months' ratings\\
\midrule
\multirow{2}{*}{0.6} & Ecc. & 3.15* & 0.13* & 45\%* \\
\cdashline{2-5}[0.5pt/1pt]
& Non-ecc.& 2.90* & 0.11* & 20\%*\\
\cmidrule(lr){2-5}
\multirow{2}{*}{0.7} & Ecc. & 3.35* & 0.12* & 25\%*\\
\cdashline{2-5}[0.5pt/1pt]
& Non-ecc.& 2.87* & 0.10* & 11\%*\\
\cmidrule(lr){2-5}
\multirow{2}{*}{0.8} & Ecc. & 2.98 & 0.41* & 84\%*\\
\cdashline{2-5}[0.5pt/1pt]
& Non-ecc.& 3.04 & 0.37* & 75\%*\\
\cmidrule(lr){2-5}
\multirow{2}{*}{0.9} & Ecc. & 2.87 & 0.40 & 82\%*\\
\cdashline{2-5}[0.5pt/1pt]
& Non-ecc.& 2.99 & 0.38 &80\%*\\
\cmidrule(lr){2-5}
\multirow{2}{*}{0.99} & Ecc. & 2.90 & 0.41 & 84\%*\\
\cdashline{2-5}[0.5pt/1pt]
& Non-ecc.& 2.99 & 0.38 & 77\%*\\
\bottomrule
\end{tabular}}
\end{table}

The previous experiments showed that there are consistently eccentric and noneccentric items, and they are mixed in the tail part.
We now investigate these different groups of items in detail and show how they are different.
We first extracted various features of items, such as an average number of users, mean feedback score, and popularity of the artist.
We then kept the first quintile (top 20\%) and the last quintile (bottom 20\%) groups of items divided by their eccentricity value, and respectively labeled them as \textit{eccentric} and \textit{noneccentric} groups.
We isolated these two IE groups on each 0.6, 0.7, 0.8, 0.9, 0.99 IR group. The $X$ IR group is defined as the set of items whose rarity value is located between plus--minus one percentile from $X$.
For example, the 0.6 IR group contains the items whose IR value locates between 59 percentile and 61 percentile.
After dividing the groups, we finally averaged several features of items in each group.
We conducted statistical tests to see whether the two eccentricity groups in the same rarity group are significantly different.
Tables \ref{tab:30group}, \ref{tab:mlgroup} present the most significant and consistent differences between the \textit{eccentric} and \textit{noneccentric} groups among the several features.

Attributes of music items presented in Table \ref{tab:30group} are the \textit{listened to once users}, \textit{\# of plays}, \textit{\# of loved users}, \textit{\# of users of artists}.
The \textit{listened to once users} indicates the percentage of users who only listened to the track once, and did not listen again. The eccentric track groups had smaller \textit{listened to once users} and had higher \textit{\# of plays} than the noneccentric groups.
This signifies that the eccentric tracks are more likely to be replayed by users in future. Higher average \textit{\# of loved users} of eccentric tracks indicates that the tracks were not only listened to by users in greater number but also were favored by people.
The last column, \textit{\# of users of artists}, was the most prominent difference between the two groups.
Noneccentric tracks were from more popular artists compared with eccentric tracks.
This implies that the major difference between eccentric and noneccentric items may lie in the artists.

For the movies, the differences between eccentric and noneccentric groups were well captured by three features:
\textit{rating}, \textit{multimodality of ratings}, and \textit{first two months' ratings}.
However, the difference between the two eccentricity groups was not as evident as the music because of the difference in the value range of feedbacks.
The users' feedback is used to give weight to their eccentric behavior and increase the ability to separate items' eccentricity value. However, the range of ratings is limited to 0.5--5 for movies while users can replay tracks as much as they want.
Therefore, in the extreme tail, the IR 0.9 and 0.99 groups of the MovieLens dataset did not show a clear difference.

Considering the significant results presented in Table \ref{tab:mlgroup}, we could observe that the eccentric movies have higher ratings and their ratings tend to follow the multimodal Gaussian distribution. \textit{Multimodality of ratings} was measured by Hartigan's Dip test statistic \cite{Hartigan}. This implies that the ratings are more polarized for eccentric and rare items.
The last column, \textit{first two months' ratings}, was the only feature where all IR groups had a significant difference between the two eccentricity groups.
The result shows that the ratings of eccentric items are more concentrated on the time they were released, while the noneccentric items were also consumed after the release.

\section{Conclusion}
\label{sec:con}
Our work has proposed a novel metric to measure the eccentricity of items.
The main intuition behind our metric was that the eccentricity of items is related to who are the consumers, not to the number of people who have used the item.
Using the proposed metric, we have analyzed two real-world datasets.
Through the analysis, we showed that the eccentric and noneccentric items classified by our metric maintained their eccentricity value regardless of the time being measured, and showed a significant difference on various features.

\section*{Acknowledgment}
This work was supported by ICT R\&D program of MSIP/IITP (R7124-16-0004, Development of Intelligent Interaction Technology Based on Context Awareness and Human Intention Understanding).

\bibliographystyle{IEEEtran}
\bibliography{SMC_2017_chan}

\end{document}